\documentclass[aps,prd,10pt,notitlepage,nofootinbib,superscriptaddress,showkeys,showpacs]{revtex4-2}
\pdfoutput=1
\usepackage{amstext,amsmath,amssymb,amsfonts,bbm,euscript,color,dsfont}
\usepackage[utf8]{inputenc}
\usepackage{epsfig}
\usepackage{hyperref}
\usepackage{slashed}
\usepackage{amsthm}
\usepackage{subfigure}
\usepackage{xcolor}
\usepackage{multirow}
\usepackage{bbold}
\newcommand{\Tr}{\operatorname{Tr}}
\usepackage{bm}
\usepackage{tikz}
\usetikzlibrary{positioning,arrows.meta}
\usepackage{verbatim}
\usepackage{graphicx}
\usepackage{latexsym}

\begin{document}

\title{Quark sector effects and glueball mass sensitivity estimates in Lorentz-violating supersymmetric QCD-like theories}

\author{Rodrigo Carmo Terin} 
\email{rodrigo.carmo@urjc.es}
\affiliation{King Juan Carlos University, Faculty of Experimental Sciences and Technology, Department of Applied Physics, Av. del Alcalde de Móstoles, 28933, Madrid, Spain}

\begin{abstract}
The $N=1$ supersymmetric Yang--Mills--Carroll--Field--Jackiw (SYM--CFJ) model is extended
to include the quark sector of supersymmetric quantum chromodynamics (SQCD) in the presence of Lorentz--symmetry
violation (LSV). The Lorentz--violating data are carried by a spurion chiral superfield whose
components define a purely spacelike background vector $v_\mu$ and fermionic bilinears,
inducing a topological mass scale in the gauge sector. Working within a
spurion effective field theory (EFT) and a small--LSV expansion, we classify the allowed parametric dependence of glueball observables on the
induced mass scales. Using lattice Yang--Mills (YM) glueball masses only as a reference hadronic scale, we provide parametric sensitivity estimates and naturalness
ranges for the topological mass $|v|$ relative to $\Lambda_{\rm YM}$.
\end{abstract}

\maketitle
\section{Introduction}
\label{sec:introduction}

Lorentz invariance and $CPT$ symmetry are central pillars of relativistic quantum
field theory and of the Standard Model. Nevertheless, it is widely recognized
that, at very high energies, for instance, near the Planck scale these
symmetries may be only emergent, being spontaneously or explicitly violated by
more fundamental dynamics, such as those contemplated in string-inspired
settings \cite{KosteleckySamuel1989a,KosteleckyPotting1991}.
At the level of low-energy phenomenology, such effects are efficiently organized
within the standard model extension (SME), where Lorentz-violating operators are
constructed by contracting Standard Model fields with fixed background tensors
\cite{ColladayKostelecky1997,ColladayKostelecky1998,KosteleckyRussell2011}.
Within this program, the gauge sector offers particularly clean probes: both
$CPT$-odd deformations, such as the CFJ term
\cite{CarrollFieldJackiw1990}, and $CPT$-even deformations, such as the
$k_F$-type operators \cite{KosteleckyMewes2002}, have been extensively studied
and tightly constrained by laboratory and astrophysical tests
\cite{KosteleckyRussell2011}.

A natural theoretical question then concerns the role of supersymmetry (SUSY) in
scenarios with LSV. If SUSY is realized at high
energies, LSV should plausibly arise in a supersymmetric environment, so that
the Lorentz-violating background must belong to some supermultiplet. This
expectation motivates supersymmetric embeddings of Lorentz-violating operators,
and several works have clarified that LSV and SUSY breaking are generically
intertwined: once the background selects a preferred direction in spacetime, the
supersymmetric partners of the Lorentz-violating data typically induce
non-trivial fermionic condensates and mass splittings
\cite{BergerKostelecky2002,NibbelinkPospelov2005,Redigolo2012}.

Moreover, supersymmetric theories with LSV can be formulated
following two complementary strategies. One viewpoint is to deform the
supersymmetry algebra itself, leading to a modified superfield formalism in
deformed superspace and typically to aether-like, CPT-even Lorentz-violating
structures \cite{Farias:2012at,LehumSupergaugeAether2013}. An alternative approach,
adopted in the present work, preserves the standard supersymmetry algebra and
introduces LSV through background superfields treated as
spurions, which naturally accommodates CPT-odd deformations, e.g., the
CFJ term in the gauge sector.

In this direction, a supersymmetric realization of the CFJ mechanism can be
implemented by introducing a background chiral superfield $S$, whose components
define a constant four-vector $v_\mu$ and fermionic bilinears. In this work we
restrict to a purely spacelike choice of $v_\mu$, thereby avoiding the well-known
unitarity and stability issues associated with timelike or lightlike CFJ
backgrounds. Therefore, throughout this work, the Lorentz-violating superfield $S$ is treated as a fixed
non-dynamical background in the standard spurion sense, so that symmetry
constraints, rather than dynamical equations of motion, determine the allowed
operators and mass deformations.

Early supersymmetric embeddings of CFJ-like terms were developed in Abelian gauge theories \cite{PhysRevD.68.065030}, notably in the Maxwell–Chern–Simons case, establishing the superspace realization of Lorentz- and CPT-violating backgrounds, albeit without addressing non-Abelian dynamics or matter sectors. In the non-Abelian case, this construction was explicitly carried out in one of our previous works~\cite{TerinSYMCS2022}, where an $\mathcal{N}=1$ super--YM
extension of the CFJ term was shown to preserve gauge invariance while inducing a
topological mass scale in the gauge sector and condensate-driven effects in the
gaugino sector, leading to characteristic modifications of the elementary
dispersion relations. Related analyses in Abelian and $CPT$-even supersymmetric
LSV scenarios further support the physical relevance of background fermion
bilinears beyond purely kinematical deformations
\cite{BelichPhotino2013,BelichCPTeven2015}.

These results suggest that LSV in a supersymmetric environment can influence both elementary spectra and the infrared dynamics of non-Abelian gauge
theories. Motivated by this perspective, a natural next step is to incorporate
matter fields and investigate how LSV manifests itself in SQCD, where gauge fields, gauginos, quarks and squarks interact
in a highly constrained setting and non-perturbative effects play a central
role. Recent progress in lattice formulations and renormalization of SQCD,
including the structure of counterterms and fine-tuning requirements
\cite{CostaPanagopoulos2017,CostaPanagopoulos2019,CostaYukawa2024,CostaQuartic2024},
opens the possibility of extending Lorentz-violating supersymmetric gauge
theories beyond the pure gauge sector in a controlled manner.

Within this supervised setting, we discuss how these deformations propagate into
the dynamics of gauge-invariant composite operators, with emphasis on glueball
channels understood as effective color-singlet excitations of the gluonic sector~\cite{AltschulSchindler2019}.
In pure-gauge $\mathrm{SU}(3)$ YM theory, lattice simulations
place the lightest scalar $0^{++}$ state near $1.6~\mathrm{GeV}$
\cite{MorningstarPeardon1999,Chen2006,AthenodorouTeper2021}.
Because no dynamical quarks are present in these simulations, this value should
be regarded as a hadronic scale emerging from pure YM dynamics rather
than as a direct prediction for SQCD. Nevertheless, it provides a natural
infrared benchmark against which additional mass scales induced by Lorentz
violation can be parametrically assessed.
From an EFT viewpoint, such a comparison should be
understood as an organization of scales and sensitivities rather than as a
first-principles computation of the spectrum
\cite{ColemanWeinberg1973,Jackiw1974}.

Our strategy is therefore to use the known glueball mass spectrum as an infrared, hadronic benchmark to assess the parametric sensitivity of the Lorentz--violating parameters.  Since the SYM--CFJ deformation already introduces a topological mass scale in the gauge sector controlled by $|v|$, requiring compatibility with the established glueball mass hierarchy suggests that $|v|$ should be small compared to the hadronic scale.  We interpret this as a naturalness or sensitivity estimate rather than a sharp experimental bound, and we shiw it through a spurion EFT analysis.  Our present work thus lays the groundwork
for a systematic non--perturbative analysis of SQCD with LSV and for future lattice--oriented implementations of the spurion deformation.

This paper is organized as follows. In Sec.~\ref{sec:SYM-CFJ}, we briefly review
the $\mathcal{N}=1$ SYM--CFJ model and summarize the gauge-boson and gaugino
dispersion relations. In Sec.~\ref{sec:quarkSector}, we introduce quark and squark
superfields in the presence of the LSV spurion and derive the corresponding
bilinear deformations. In Sec.~\ref{sec:glueballs}, we discuss glueball operators
and the use of lattice benchmarks to constrain the Lorentz-violating parameters.
We conclude and present future perspectives in Sec.~\ref{sec:conclusion}.
\section{A brief overview of the SYM--CFJ model\label{sec:SYM-CFJ}}

We begin by briefly reviewing the construction of the SYM--CFJ model introduced in our previous work~\cite{TerinSYMCS2022}.  Consider an $N=1$ supersymmetric gauge theory with gauge group $G$, described in superspace by the vector superfield $V=V^a T^a$ in Wess--Zumino gauge and the chiral superfield $S$ carrying no gauge indices.  The standard pure SYM action is
\begin{equation}
  S_{\mathrm{SYM}} = \frac{1}{g^2} \int\!d^4x\,d^2\theta\, \mathrm{Tr}(W^\alpha W_\alpha) + \mathrm{h.c.},
\end{equation}
where $W_\alpha = \bar{D}^2 (\mathrm{e}^{-V} D_\alpha \mathrm{e}^V)$ is the gauge--covariant field strength superfield and $g$ is the gauge coupling.
The CFJ term arises by coupling the spinor superfield $W_\alpha$ to $S$:
\begin{equation}
  S_{\mathrm{CFJ}} = \frac{1}{g} \int\!d^4x\,d^2\theta\,d^2\bar{\theta}\;\mathrm{Tr}\big[W^\alpha \Gamma_\alpha S + \bar{W}_{\dot{\alpha}} \bar{\Gamma}^{\dot{\alpha}} \bar{S}\big],
\end{equation}
in which $\Gamma_\alpha$ is the gauge connection superfield defined by $\mathrm{e}^{-V} D_\alpha \mathrm{e}^V = D_\alpha + \Gamma_\alpha$.  Upon expanding in component fields and removing auxiliaries, the CFJ term reduces to
\begin{equation}
  S_{\mathrm{CFJ}} = \int\!d^4x\;\Big[ -\tfrac{1}{2}(s+s^*)F^a_{\mu\nu}F^{\mu\nu a} + \epsilon^{\mu\nu\rho\sigma} v_\mu A^a_\nu \partial_\rho A^a_\sigma - \tfrac{g}{3}\epsilon^{\mu\nu\rho\sigma} f^{abc} v_\mu A^a_\nu A^b_\rho A^c_\sigma + \cdots\Big],
\end{equation}
where $s$ and $\psi$ are the scalar and fermion components of $S$, $v_\mu = \partial_\mu (\mathrm{Im}\,s)$ is the background four--vector, and the ellipsis denotes gaugino mass terms and fermion bilinears. At this point, let us comment on the role of the real part of the scalar
background. If \(\mathrm{Re}\,s\) is taken to be constant, the first term in
Eq.~(3) amounts only to a finite rescaling of the gauge kinetic term and can be
absorbed into a redefinition of the gauge-field normalization and of the gauge
coupling. Alternatively, one may consistently choose the background
\(\mathrm{Re}\,s=0\). In either case, this term does not affect the
Lorentz-violating effects discussed below, which are controlled by the CFJ
background
\[
v_\mu=\partial_\mu \mathrm{Im}\,s .
\]
We shall therefore focus on the CFJ deformation and on its supersymmetric
fermionic partners. Gauge invariance demands that $v_\mu$ be curl--free; supersymmetry further fixes it to be a derivative of the scalar field.
From the component action one can derive the equations of motion and linearizing around the background and using plane--wave solutions yields dispersion relations for the gauge boson $A^a_\mu$ and gaugino $\lambda^a$.  The gauge boson satisfies
\begin{equation}
  (k^2 \eta_{\mu\nu} - k_\mu k_\nu + i \epsilon_{\mu\nu\rho\sigma} v^\rho k^\sigma)A^{\nu a} = 0,
\end{equation}
whose determinant leads to the characteristic equation $\omega(\omega^2 - |v|^2)=0$. Here \(k^\mu=(\omega,\mathbf{k})\), with \(\omega\) denoting the energy of the
gauge mode. The behavior of the CFJ dispersion relation depends on the causal
character of \(v_\mu\). Timelike and lightlike backgrounds are known to be
associated with stability and/or unitarity subtleties, while a purely spacelike
background gives a better behaved spectrum. For this reason, in the present work
we restrict to
\[
v_\mu=(0,\mathbf v).
\]. Linearizing the gauge-field equations in the CFJ background yields modified
dispersion relations with a topological mass scale set by the spacelike vector,
\begin{equation}
  m_A \sim |v| ,
\end{equation}
up to convention-dependent numerical factors and polarization-dependent
details that are not needed for the present spurion-level analysis. Thus the gauge boson acquires a topological mass scale
$m_A\sim|v|$, which will play the role of the primary infrared deformation
parameter in the subsequent spurion EFT analysis.

The gaugino obeys a modified Dirac equation with axial couplings to $v_\mu$ and the fermionic bilinear $\Psi\Psi$, where $\Psi$ is the fermion in $S$.  In momentum space at rest ($\bm{k}=0$), one finds the dispersion relation
\begin{equation}
\det\Big[\omega \gamma^0 + \tfrac{1}{2} \big(\langle\Psi\Psi\rangle \pm |\bm{v}|\big)\mathbb{1}_4\Big]=0,
\end{equation}
Our solutions of this dispersion relation depend schematically on the background bilinear $\langle\Psi\Psi\rangle$ and on $|\bm{v}|$.  Because $\langle\Psi\Psi\rangle$ has mass dimension three whereas $|\bm{v}|$ has dimension one, the overall normalization of these solutions is model dependent and tied to the ultraviolet completion.  We therefore refrain from identifying these quantities with literal mass eigenvalues and instead view the resulting splitting as an indicative signal of supersymmetry breaking induced by LSV. Since $\gamma^0$ has eigenvalues $\pm 1$, each with degeneracy two, a determinant
of the form
\[
\det(\omega\gamma^0+M\mathbbm{1}_4)
\]
reduces to
\[
\det(\omega\gamma^0+M\mathbbm{1}_4)=(M^2-\omega^2)^2.
\]
Thus, for each effective background value \(M_\pm\), the dispersion condition is
\[
\omega^2=M_\pm^2 .
\]
In the present EFT treatment, however, \(M_\pm\) should be understood as a
background-induced effective parameter whose normalization depends on the
UV completion.
\section{Quark Sector with LSV\label{sec:quarkSector}}

To extend the SYM--CFJ construction to SUSY--QCD, we add chiral superfields
$\Phi_i$ (quark flavours) transforming in a representation $R$ of $G$ and their
conjugates $\bar\Phi^{\,i}$ in $\bar R$,
\begin{equation}
  \Phi_i' = e^{i\Xi^a T^a_R}\,\Phi_i,\qquad
  \bar\Phi^{\,i\prime} = \bar\Phi^{\,j}\,e^{-i\bar\Xi^a T^a_R}
  \,
\end{equation}
where $\Xi$ is chiral. The standard
gauge-invariant matter action reads
\begin{equation}
  S_{\rm matter}^{(0)}
  = \sum_i \int d^4x\,d^4\theta\;\bar\Phi^{\,i}\,e^{2gV}\,\Phi_i
  +\left(\int d^4x\,d^2\theta\;W(\Phi_i)+{\rm h.c.}\right),
\end{equation}
with $W$ the superpotential. In components (Wess--Zumino gauge),
\begin{align}
  \Phi_i &= \phi_i + \sqrt2\,\theta\,\psi_i + \theta^2 F_i,\nonumber\\
  V &= -\theta\sigma^\mu\bar\theta\,A_\mu
  + i\theta^2\bar\theta\,\bar\lambda
  - i\bar\theta^2\theta\,\lambda
  +\frac12\theta^2\bar\theta^2 D \,.
\end{align}
The fermion $\psi_i$ is a left-handed Weyl quark; together with its conjugate it
builds a Dirac quark $q_i$. The gauge and Yukawa interactions are standard and
will only matter here insofar as they constrain which Lorentz-violating operators
are allowed.

In our SYM--CFJ setup, the Lorentz-violating data are not inserted ``by hand'' as
arbitrary tensors: they descend from the background chiral superfield $S$,
\begin{equation}
  S = s + \sqrt2\,\theta\psi_S + \theta^2 F_S
  + i\theta\sigma^\mu\bar\theta\,\partial_\mu s + \cdots,
\end{equation}
whose components yield the constant CFJ vector and fermion bilinears (in the
vacuum configuration used in the gauge/gaugino dispersion section):
\begin{equation}
  v_\mu \equiv \partial_\mu\,\mathrm{Im}\,s \qquad(\mathrm{taken\ as\ constant}),\qquad
  \Psi \equiv \psi_S,\qquad
  F \equiv F_S.
\end{equation}
Treating $S$ as a spurion (a background superfield with fixed expectation
values) is the clean way to classify all gauge-invariant, superspace operators
that:
\begin{enumerate}
  \item[(i)] reduce to Lorentz-violating terms in components,
  \item[(ii)] respect gauge symmetry, and
  \item[(iii)] are suppressed by some heavy scale $M$ (the scale at which the
  microscopic LSV dynamics lives).
\end{enumerate}
Dimensional analysis:
\[
[d^4\theta]=2,\quad [d^2\theta]=1,\quad [\Phi]=1,\quad [V]=0,\quad [S]=1.
\]
Thus, the leading gauge-invariant couplings of $S$ to matter can be organized
as higher-dimensional superspace operators suppressed by a heavy scale $M$.
For later use, it is convenient to distinguish: (i) CPT-even operators that can
generate Lorentz-invariant mass-type deformations once the spurion acquires
VEVs, and (ii) CPT-odd derivative operators that reduce to SME-like couplings
controlled by $v_\mu \equiv \partial_\mu \mathrm{Im}\,s$.

A convenient (non-unique) operator basis capturing the effects needed for our
parametric estimates is
\begin{align}
  \Delta S_{\rm matter}^{\rm even}
  &=
  \frac{c_K}{M}\sum_i\int d^4x\,d^4\theta\;
  \bar\Phi^{\,i}e^{2gV}\Phi_i\,(S+S^\dagger)
  \label{eq:dim5Kahler}
  \\
  &\quad
  +\frac{c_{SS}}{M^2}\sum_i\int d^4x\,d^4\theta\;
  (S^\dagger S)\,\bar\Phi^{\,i}e^{2gV}\Phi_i
  \label{eq:dim6SSKahler}
  \\
  &\quad
  +\frac{1}{M}\sum_i\left[
  \int d^4x\,d^2\theta\; S\,m_{ij}\,\Phi_i\Phi_j
  +{\rm h.c.}\right],
  \label{eq:dim5superpot}
  \\
  \Delta S_{\rm matter}^{\rm odd}
  &=
  \frac{c_D}{M^2}\sum_i\int d^4x\,d^4\theta\;
  \bar\Phi^{\,i}e^{2gV}\,\Big(\overleftrightarrow{\nabla}_\mu\Phi_i\Big)\,
  \partial^\mu(S-S^\dagger)
  +\cdots,
  \label{eq:dim6CPTodd}
\end{align}
where $c_K,c_{SS},c_D$ are dimensionless Wilson coefficients expected to be
$\mathcal{O}(1)$ in the absence of symmetries. The operator
\eqref{eq:dim5Kahler} mainly induces a spurion-dependent wavefunction
renormalization and scalar/auxiliary mixings, while the genuinely mass-type
CPT-even deformations are most cleanly represented by the superpotential operator
\eqref{eq:dim5superpot} and/or the quadratic spurion K\"ahler operator
\eqref{eq:dim6SSKahler}. The operators in Eqs.~\ref{eq:dim5Kahler}--\ref{eq:dim6CPTodd} are higher-dimensional interactions and,
therefore, the matter-spurion sector should not be interpreted as a power-counting
renormalizable completion of SQCD. Rather, it is an effective field theory
valid below the scale \(M\). In this sense, the non-renormalizable terms represent
the leading low-energy imprint of the Lorentz-violating supersymmetric
background.

We now show explicitly (at the level needed for estimates) how the terms in
\eqref{eq:dim5Kahler}--\eqref{eq:dim6CPTodd} generate the effective operators
in the quark/squark sector. The logic is the same as in the gauge/gaugino
dispersion section: one isolates the bilinear terms and studies their impact on
free propagation.

\paragraph{CPT-even mass-type deformations.}
Consider first the K\"ahler-type operator \eqref{eq:dim5Kahler}. For a
constant spurion scalar background $(s+s^\dagger)=\mathrm{const.}$, its leading
effect at the bilinear level is a spurion-dependent rescaling of the matter
kinetic terms (hence of the wavefunction normalization). It also generates
mixings involving the auxiliary field $F_i$ and the spurion auxiliary $F_S$,
which may feed into scalar mass terms once auxiliaries are eliminated and a
superpotential is specified. For genuinely mass-type CPT-even deformations that
are manifest already at the spurion level, it is cleaner to use the
superpotential operator \eqref{eq:dim5superpot} and/or the quadratic-spurion
K\"ahler operator \eqref{eq:dim6SSKahler}.

From \eqref{eq:dim5superpot}, writing $S=\langle s\rangle + \theta^2\langle F_S\rangle + \cdots$,
one obtains the standard component-level bilinears
\begin{equation}
  \Delta\mathcal{L}_{\rm even}
  \supset
  -\frac{1}{M}\,m_{ij}\,\langle s\rangle\;\psi_i\psi_j
  -\frac{1}{M}\,m_{ij}\,\langle F_S\rangle\;\phi_i\phi_j
  +{\rm h.c.},
\end{equation}
which induce a SUSY-preserving fermion mass (controlled by $\langle s\rangle$)
and a holomorphic scalar mass term (managed by $\langle F_S\rangle$) in the
presence of SUSY breaking. In addition, the operator \eqref{eq:dim6SSKahler}
generates CPT-even, Lorentz-invariant bilinears proportional to spurion
invariants such as $\langle S^\dagger S\rangle$, which may be viewed
parametrically as encoding contributions from backgrounds like $\bar\Psi\Psi$.
Accordingly, at the level needed for estimates we package CPT-even mass-type
effects as
\begin{equation}
  \delta m_q \sim \frac{1}{M}\,m\,\langle s\rangle
  \;+\;\frac{1}{M^2}\,\kappa\,\langle S^\dagger S\rangle,
  \qquad
  \delta m_{\tilde q}^2\ \ \text{analogous (and SUSY-breaking sensitive)}\,,
  \label{eq:delta_mq_package}
\end{equation}
where $\kappa$ denotes an $\mathcal{O}(1)$ Wilson coefficient and the precise
mapping between $\langle S^\dagger S\rangle$ and fermion bilinears depends on the
spurion sector and UV completion.

\paragraph{CPT-odd derivative coupling and SME-like currents.}
Now consider the derivative operator \eqref{eq:dim6CPTodd} and insert
\begin{equation}
  \partial_\mu(S-S^\dagger)=2i\,\partial_\mu\mathrm{Im}\,s \equiv 2i\,v_\mu,
  \qquad v_\mu=\mathrm{const.}
\end{equation}
At the component level, the lowest terms in
$\bar\Phi e^{2gV}\overleftrightarrow{\nabla}_\mu\Phi$ contain the standard
matter currents. Keeping only bilinears, one obtains schematically
\begin{equation}
  \Delta\mathcal{L}_{\rm odd}
  \supset
  \frac{c_D}{M^2}\,v_\mu\,
  \Big[
    \bar\psi\,\bar\sigma^\mu\psi
    + i\big(\phi^\dagger D^\mu\phi - (D^\mu\phi)^\dagger\phi\big)
  \Big]
  +\cdots,
  \label{eq:current_component}
\end{equation}
which is the superspace origin of an SME-like coupling $v_\mu J^\mu$.
When assembling a four-component Dirac quark from two chiral multiplets (as in
SQCD with quarks and antiquarks), the induced $v_\mu J^\mu$ term can be
decomposed into vector and axial combinations depending on how the operator
acts on the left- and right-chiral multiplets. For the present spurion-level
discussion, it suffices to parameterize the CPT-odd matter deformation as
\begin{equation}
  \mathcal{L}_q^{\rm eff}
  \supset
  - v_\mu\,\bar q\,
  \big(a^\mu + b^\mu\gamma_5\big)\,q,
  \qquad
  a^\mu,\ b^\mu \sim \mathcal{O}\!\left(\frac{1}{M^2}\right),
  \label{eq:SME_ab}
\end{equation}
with $a^\mu$ and $b^\mu$ fixed by the chiral assignment (and both ultimately
controlled by $v_\mu$ and Wilson coefficients).

Here we restrict to a purely spacelike background
$v_\mu=(0,\bm v)$, so that $v^2=-\bm v^{\,2}$ and $v_0=0$.
Accordingly, at $\bm p=\bm 0$ there is no linear rest-energy splitting
proportional to $v_0$, and the leading effects of the CPT-odd term on rest
energies start at $\mathcal{O}(\bm v^{\,2})$, while direction-dependent effects
appear already at $\mathcal{O}(\bm v\!\cdot\!\bm p)$ for $\bm p\neq 0$.
We therefore package the CPT-odd quark deformation as
\begin{equation}
  a_\mu \sim \frac{\beta_a}{M^2}v_\mu,
  \qquad
  b_\mu \sim \frac{\beta_b}{M^2}v_\mu,
  \qquad
  v_\mu=(0,\bm v),
  \label{eq:ab_package}
\end{equation}
with $\beta_a,\beta_b=\mathcal O(1)$. In this spacelike setup, the cleanest
leading probes of Lorentz violation are thus anisotropies and multiplet
splittings, whereas purely rest-mass effects typically start only at
$\mathcal{O}(\bm v^{\,2})$ unless additional symmetry-breaking sources are present. Indeed, this observation has a direct impact on the glueball analysis developed in
Sec.~\ref{sec:glueballs}. Since purely rest-mass shifts are suppressed for a
spacelike background, the most robust signatures of Lorentz violation in the
gluonic sector are expected to arise from anisotropic dispersion relations,
splittings of rotational multiplets, and channel-dependent sensitivities to the
induced gauge-sector mass scale $m_A\sim|\bm v|$. These features provide
parametrically clean probes that can be confronted with lattice glueball
spectroscopy without relying on a first-principles computation of the spectrum.

Collecting the leading bilinear terms for a given flavour (and turning off gauge
fields for the dispersion estimate, as in the gaugino sector), the effective quark
Lagrangian takes the SME-like form
\begin{equation}
  \mathcal{L}_q^{\rm eff}
  = \bar q\Big(i\slashed{\partial} - m_q^0 - \delta m_q
  - \slashed{b}\,\gamma_5\Big)q + \cdots,
  \qquad
  b_\mu \equiv \frac{\beta}{M^2}v_\mu ,
  \label{eq:SMEform_clean}
\end{equation}
with $\delta m_q$ generated by CPT-even spurion operators and $b_\mu$ by the
CPT-odd derivative coupling.

For the purely spacelike background adopted throughout this work,
$v_\mu=(0,\bm v)$, one has $b_0=0$. Accordingly, at vanishing spatial momentum
the quark rest energy receives no linear correction from the CPT-odd term.
Instead, the leading modification is quadratic,
\begin{equation}
  E(\bm p=\bm 0)
  = \sqrt{m^2+\bm b^{\,2}}
  \;\simeq\;
  m + \frac{\bm b^{\,2}}{2m}
  \;=\;
  m + \mathcal{O}\!\left(\frac{\bm v^{\,2}}{M^4}\right),
\end{equation}
while direction-dependent effects appear already at
$\mathcal{O}(\bm v\!\cdot\!\bm p)$ for $\bm p\neq 0$.

Therefore, in the spacelike CFJ regime relevant for our analysis, the effective
quark mass entering infrared observables can be consistently parameterized as
\begin{equation}
  m_q^{\rm eff}
  \simeq
  m_q^0
  + \delta m_q
  + \mathcal{O}\!\left(\frac{\bm v^{\,2}}{M^4}\right),
  \qquad
  \delta m_q
  \sim
  \frac{\alpha}{M}\Big(\mathrm{Re}\,F + \tfrac14\bar\Psi\Psi\Big),
  \label{eq:mqeff_spacelike}
\end{equation}
with no linear rest-energy splitting induced by $v_\mu$.

The same spurion operators also induce corrections in the squark sector.
Keeping only bilinears and denoting by $\tilde q$ a generic squark field, the
effective Lagrangian reads
\begin{equation}
  \mathcal{L}_{\tilde q}^{\rm eff}
  =
  |D_\mu\tilde q|^2
  - (m_{\tilde q}^0)^2\,\tilde q^\dagger\tilde q
  - \delta m_{\tilde q}^2\,\tilde q^\dagger\tilde q
  + \frac{c_{\tilde q}}{M^2}v_\mu\,
  i\big(\tilde q^\dagger D^\mu\tilde q - (D^\mu\tilde q)^\dagger\tilde q\big)
  +\cdots,
\end{equation}
with
\begin{equation}
  \delta m_{\tilde q}^2 \sim
  \frac{1}{M}\Big(\mathrm{Re}\,F + \tfrac14\bar\Psi\Psi\Big)\Lambda_{\rm soft}
  + \cdots,
\end{equation}
where $\Lambda_{\rm soft}$ indicates the sensitivity of scalar masses to the
SUSY-breaking scale once the background spurion is switched on.

With dynamical quarks, two facts become fundamental for the glueball programme.
First, glueball operators mix with $q\bar q$ operators (in particular in the
scalar channel), and their correlators acquire additional intermediate states.
The LSV-induced shifts in the effective quark and squark parameters
move (a) quark thresholds,
(b) the location of branch cuts in correlators, and
(c) the amount of mixing through modified propagators.
Accordingly, even small spurion backgrounds $(\bm v,\bar\Psi\Psi)$ can lead to
parametrically clean effects in glueball channels once compared to lattice
spectra. Second, Lorentz violation introduces preferred directions, therefore rotational
multiplets can split and glueball dispersions can become anisotropic; this makes
ratios and splittings of extracted masses a robust way to constrain the same LSV
spurion parameters that also control the topological gauge-sector mass. For later matching it is useful to package the matter-sector LSV data as
\begin{equation}
  \delta m_q \sim \frac{\alpha}{M}\Big(\mathrm{Re}\,F + \tfrac14\bar\Psi\Psi\Big),
  \qquad
  b_\mu \sim \frac{\beta}{M^2}v_\mu,
  \qquad
  \delta m_{\tilde q}^2 \ \text{analogous but SUSY-breaking sensitive}.
\end{equation}
These are the quantities that enter quark and squark propagators inside glueball
and hybrid correlators and therefore determine how the lattice-extracted
glueball spectrum constrains the fundamental LSV data encoded in the spurion $S$.

Although the quark sector does not provide a direct determination of glueball
masses, it is fundamental in the present analysis. All
Lorentz-violating effects entering the matter sector are controlled by the same
spurion superfield $S$ that induces the CFJ deformation in the gauge sector.
As a consequence, quark and squark propagators acquire only spurion-suppressed
deformations, characterized by $\delta m_q$ and $b_\mu$, which modify glueball
correlators indirectly through loop effects, mixing with $q\bar q$ operators,
and threshold shifts. The purpose of the quark-sector analysis is to ensure the internal consistency of the EFT
description and to constrain the parametric structure and natural size of the
Lorentz-violating corrections entering gauge-invariant composite observables.
\section{Glueball sector in the SYM--CFJ--SQCD model}
\label{sec:glueballs}

Glueballs are colour--singlet bound states built predominantly from gluonic
degrees of freedom. In Euclidean lattice Yang--Mills theory, glueball masses are
extracted from temporal correlation functions of gauge--invariant composite
operators with fixed $J^{PC}$ quantum numbers.

Before proceeding, we stress that, in the absence of a fully nonperturbative
solution of the SYM--CFJ--SQCD theory, the glueball spectrum discussed in this
section is not computed from first principles. Instead, we derive the most
general parametric dependence of glueball observables on the Lorentz--violating
background, guided by symmetry, dimensional analysis, and by the induced
gauge--sector mass scale.  
Accordingly, the present analysis is performed in a spurion EFT sense: the
background superfield is treated as non-dynamical, and we classify the allowed
functional dependence of correlators on $v_\mu$ and on the fermionic bilinears
rather than computing glueball masses explicitly. Our strategy parallels the SME-based EFT treatment of hadrons, where Lorentz-violating quark and gluon operators are matched onto composite observables without a first-principles computation of the spectrum, see e.g.~\cite{AltschulSchindler2019}
A complete quantization of the model, including the implementation of BRST
symmetry, Slavnov--Taylor identities and a systematic renormalization analysis in
superspace, is left for future work.

A standard lattice strategy is to construct a variational basis
$\{\mathcal O_i^{(J^{PC})}\}$ of gauge--invariant operators (Wilson loops of
various shapes, clover discretizations of $F_{\mu\nu}$, smeared operators, etc.)
and to form the correlator matrix
\begin{equation}
  C_{ij}(t)\equiv
  \langle\, \mathcal O_i(t)\,\mathcal O_j^\dagger(0)\,\rangle,
  \qquad t>0,
\end{equation}
whose spectral decomposition reads
\begin{equation}
  C_{ij}(t)=\sum_{n} Z_i^{(n)} Z_j^{(n)\,*} e^{-E_n t},
  \qquad
  Z_i^{(n)}\equiv \langle 0|\mathcal O_i|n\rangle .
\end{equation}
The stationary energies $E_n$ are extracted from the generalized eigenvalue
problem (GEVP),
\begin{equation}
  C(t)\,v^{(n)}(t,t_0)=\lambda^{(n)}(t,t_0)\,C(t_0)\,v^{(n)}(t,t_0),
\end{equation}
for which the principal correlators behave asymptotically as
\begin{equation}
  \lambda^{(n)}(t,t_0)
  = e^{-E_n (t-t_0)}
  \left[1+\mathcal O\!\left(e^{-(E_{n+1}-E_n)(t-t_0)}\right)\right].
\end{equation}
In pure $\mathrm{SU}(3)$ Yang--Mills theory, lattice simulations locate the
lightest scalar glueball $0^{++}$ around
$m_{0^{++}}^{(0)} \simeq 1.5$--$1.7~\mathrm{GeV}$.
In the present work, this value is used solely as a hadronic reference scale,
not as a prediction for SQCD, but rather as an infrared benchmark to organize the
expected hierarchy of Lorentz--violating deformations.

In the SYM--CFJ framework, Lorentz violation enters through the background chiral
superfield $S$ and its components, notably the constant CFJ vector
$v_\mu \equiv \partial_\mu \mathrm{Im}\,s$, which we take to be purely spacelike,
\begin{equation}
  v_\mu=(0,\bm v), \qquad v^2=-\bm v^{\,2}.
\end{equation}
At the linearized level, the CFJ deformation induces a topological gauge--sector
mass scale
\begin{equation}
  m_A \sim |\bm v| ,
\end{equation}
already visible in the elementary gauge--boson dispersion relations reviewed in
Sec.~\ref{sec:SYM-CFJ}.

Glueball observables can therefore depend on the LSV spurion through:
\begin{enumerate}
  \item[(i)] direct modifications of gluon propagation and interaction vertices,
  \item[(ii)] indirect effects through matter (quark/squark/gaugino) loops, whose
  propagators are modified as discussed in Sec.~\ref{sec:quarkSector},
  \item[(iii)] kinematical anisotropies, since $v_\mu$ breaks rotational symmetry.
\end{enumerate}
Glueball interpolators must remain gauge invariant. A convenient continuum basis
for the lowest channels is
\begin{align}
  \mathcal O_{0^{++}}^{(0)} &= \Tr(F_{\mu\nu}F_{\mu\nu}),\\
  \mathcal O_{0^{-+}}^{(0)} &= \Tr(F_{\mu\nu}\tilde F_{\mu\nu}),\\
  \mathcal O_{2^{++},\mu\nu}^{(0)} &=
  \Tr(F_{\mu\alpha}F_{\nu\alpha})
  -\frac14\delta_{\mu\nu}\Tr(F_{\alpha\beta}F_{\alpha\beta}).
\end{align}
With the spurion $v_\mu$, new independent gauge--invariant structures appear at
the same canonical dimension, for instance
\begin{equation}
  v_\mu v_\nu\,\Tr(F_{\mu\alpha}F_{\nu\alpha}),
\end{equation}
as well as operators involving covariant derivatives. The structure in Eq.~(37) is the gauge-invariant CPT-even aether-like
Lorentz-violating operator built from the fixed background vector and the
non-Abelian field strength. Aether-like terms have been widely discussed in
Lorentz-violating field theories, including compactification scenarios
\cite{CarrollTam2008}, radiative generation of CPT-even aether-like actions
\cite{GomesAether2009}, and the explicit construction of the non-Abelian
aether-like term in four dimensions \cite{CarvalhoNonAbelianAether2019}.. On the lattice, these
correspond to an enlarged variational basis built from anisotropic combinations
of Wilson loops and clover operators.

Let $H=H_0+\delta H$ denote the Hamiltonian (or Euclidean transfer operator),
where $\delta H$ encodes the CFJ deformation and its supersymmetric companions.
For an isolated glueball state $|G\rangle$, the leading mass shift is
\begin{equation}
  \delta m_G
  = \langle G|\delta H|G\rangle ,
\end{equation}
which corresponds, in the path--integral language, to an insertion of the LSV
operator in the glueball two--point function.
For a purely spacelike background, the rest mass admits an analytic spurion
expansion of the form
\begin{equation}
  m_G(\bm v)
  = m_G^{(0)}
  + \chi_G\,\alpha_G |\bm v|
  + \gamma_G \frac{\bm v^{\,2}}{\Lambda_{\rm YM}}
  + \beta_G \frac{\bar\Psi\Psi}{M}
  + \cdots ,
  \label{eq:mG_spurion}
\end{equation}
where $\chi_G\in\{0,1\}$ encodes whether the linear term is allowed by residual
discrete symmetries in the chosen $J^{PC}$ channel.

Requiring that the LSV--induced shift in the lightest scalar channel remains below
a hadronic tolerance $\Delta\sim 0.1~\mathrm{GeV}$ leads to
\begin{equation}
  |\delta m_{0^{++}}|
  \simeq |\alpha_{0^{++}}|\,|\bm v|
  \;\lesssim\; \Delta
  \quad\Rightarrow\quad
  |\bm v|\lesssim \frac{\Delta}{|\alpha_{0^{++}}|}.
  \label{eq:v_naturalness}
\end{equation}
\begin{figure}
  \centering
  \includegraphics[width=0.45\linewidth]{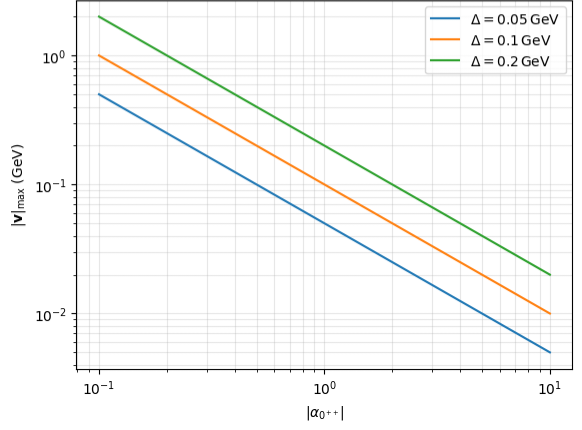}
  \caption{Naturalness estimate for the spacelike CFJ background magnitude
  $|\bm v|$ as a function of the glueball response coefficient
  $\alpha_{0^{++}}$, obtained from Eq.~\eqref{eq:v_naturalness} for representative
  tolerances $\Delta=0.05,\,0.1,\,0.2~\mathrm{GeV}$.}
  \label{fig:vmax_alpha}
\end{figure}
\newpage
The relative importance of the linear and quadratic terms in
Eq.~\eqref{eq:mG_spurion} depends on symmetry and on the magnitude of $|\bm v|$.

\begin{figure}
  \centering
  \includegraphics[width=0.45\linewidth]{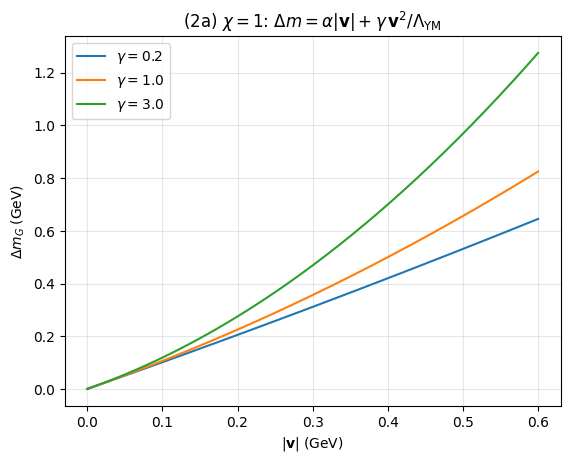}\hfill
  \includegraphics[width=0.45\linewidth]{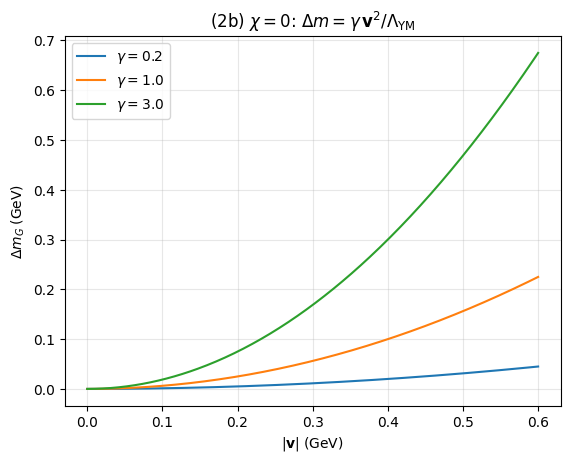}
  \caption{Theoretical behavior of the glueball mass shift
  $\Delta m_G(|\bm v|)$.
  Left: $\chi_G=1$, linear term allowed.
  Right: $\chi_G=0$, leading correction quadratic.}
  \label{fig:delta_m_v}
\end{figure}

The crossover scale between the two regimes is
\begin{equation}
  |\bm v|_\star \sim \frac{\alpha_G}{\gamma_G}\,\Lambda_{\rm YM}.
\end{equation}

\begin{figure}
  \centering
  \includegraphics[width=0.45\linewidth]{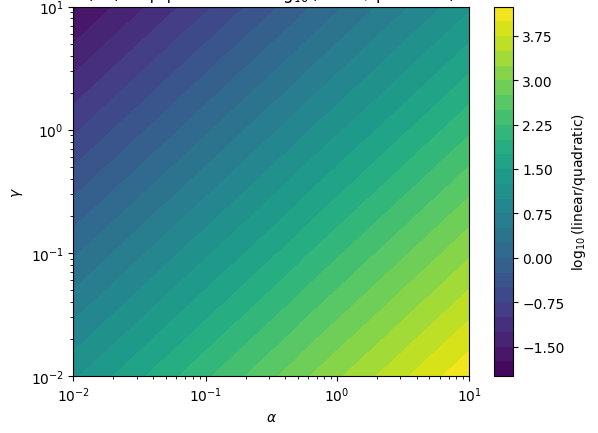}
  \caption{Crossover between linear and quadratic LSV contributions as a function
  of the EFT coefficients $(\alpha_G,\gamma_G)$.}
  \label{fig:crossover}
\end{figure}
Beyond rest masses, Lorentz violation induces anisotropic glueball dispersion
relations,
\begin{equation}
  E_G^2(\bm p)
  = m_G^2 + \bm p^{\,2}
  + \eta_G (\bm v\!\cdot\!\bm p)^2 + \cdots .
\end{equation}
This form is analogous to the anisotropic dispersion relations obtained in
aether-like supersymmetric Lorentz-breaking models\cite{Farias:2012at}, in which the preferred
background vector induces corrections proportional to contractions such as
\((v\cdot p)^2\). In the present work the same structure appears at the level of
glueball observables through the spurion expansion rather than from an explicit
first-principles solution of the bound-state problem.

\begin{figure}
  \centering
  \includegraphics[width=0.45\linewidth]{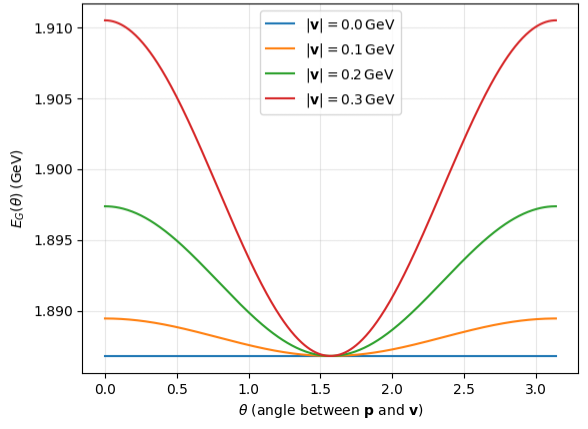}
  \caption{Angular dependence of the glueball energy $E_G(\bm p)$ at fixed
  $|\bm p|$, illustrating the anisotropy induced by a spacelike CFJ background.}
  \label{fig:anisotropy}
\end{figure}
Such anisotropies and rotational multiplet splittings provide particularly clean
lattice probes of Lorentz violation, as they are insensitive to overall mass
renormalizations.

\newpage

\section{Conclusions\label{sec:conclusion}}

In this work we have extended the super--Yang--Mills--Carroll--Field--Jackiw approach to include a SUSY--QCD matter sector and we have defined a nonperturbative strategy to assess the parametric sensitivity of LSV through
glueball spectroscopy. In the SYM--CFJ construction, LSV is encoded in a
background chiral superfield $S$ whose component $v_\mu\equiv\partial_\mu
\mathrm{Im}\,s$ selects a preferred direction in spacetime and, in the vacuum
configuration required for $v_\mu$ to be constant, supersymmetry is
simultaneously (spontaneously) broken. As a result, the gauge sector exhibits a
topological mass scale controlled by $v_\mu$, while the fermionic sector is
sensitive also to background bilinears (such as $\bar\Psi\Psi$), leading to
characteristic mass splittings between gauge bosons and gauginos.
Throughout our work we have adopted a purely spacelike choice for $v_\mu$ in order to ensure unitarity and avoid the pathologies associated with timelike or lightlike CFJ backgrounds.

We then incorporated quark and squark superfields and showed how gauge
invariance sharply restricts the admissible LSV couplings: the background enters
the matter sector through higher-dimensional, spurion-controlled operators,
which induce CPT-even mass shifts and CPT-odd axial couplings suppressed by the
microscopic LSV scale. This gives a systematic parametrization of the
low-energy LSV data in terms of effective quantities (e.g.\ $\delta m_q$ and
$b_\mu$) that feed directly into matter loops and mixing patterns in hadronic
correlators.

Motivated by the fact that glueballs are extracted from gauge-invariant Euclidean
correlators, we emphasized that the correct implementation of LSV in glueball
physics is twofold: (i) the dynamics is modified by the CFJ deformation of the
action, and (ii) the variational operator basis is enlarged by new
gauge-invariant contractions involving $v_\mu$ (rather than by
non-gauge-invariant insertions of $A_\mu$). Within this framework, glueball
masses admit an analytic small-LSV expansion, in which the induced gauge-sector
mass scale $m_A\sim|\bm v|$ provides the natural dimensionful parameter. This
makes it possible to confront the expected shifts and anisotropies in glueball
energies with lattice determinations of the pure-gauge spectrum where the
lightest scalar $0^{++}$ state lies around $1.6\,\mathrm{GeV}$
and thereby translate spectroscopy into parametric sensitivity estimates for LSV parameters rather than hard constraints.

At the level of an illustrative naturalness estimate, requiring that LSV-induced
corrections remain below the $\sim 0.1\,\mathrm{GeV}$ scale in the $0^{++}$
channel suggests a parametric relation
$|\bm v|\lesssim 0.1\,\mathrm{GeV}/|\alpha_{0^{++}}|$, where $\alpha_{0^{++}}$ is a
nonperturbative coefficient denoting the response of the scalar glueball to the
CFJ deformation. Similar (and potentially stronger) estimates can be obtained
by combining several channels, multiplet splittings, and anisotropic dispersion
tests.

Future works will therefore focus on: \begin{enumerate}
\item[(i)] quantizing the SUSY--QCD--CFJ theory and establishing the
renormalization of the LSV spurion sector (including the radiative stability of
the induced mass scales);
\item[(ii)] computing loop-induced effects of the matter sector on glueball
correlators and on glueball--meson mixing patterns in LSV backgrounds.
\item[(iii)] the use of machine-learning (ML) algorithms,
in particular physics-informed neural networks \cite{Terin10.21468,CarmoTerin:2025pvs}, to study the response of nonperturbative
correlation functions to Lorentz-violating spurions. Since the background $v_\mu$ enters
as a continuous deformation parameter, ML-based surrogate models provide a natural
framework to explore sensitivity, anisotropies and mixing effects in glueball and
hybrid channels beyond the reach of direct lattice simulations.
\end{enumerate}
These developments would turn glueball spectroscopy into a sharp probe of
Lorentz-violating supersymmetric gauge dynamics and furnish novel hadronic-scale
guidance on LSV parameters complementary to existing constraints in the standard
model extension.
\bibliographystyle{unsrt}
\bibliography{bibliography}

\end{document}